\documentclass[aps, pra, twocolumn, superscriptaddress]{revtex4-1}
\usepackage[a4paper]{geometry}
\usepackage{amsmath}
\usepackage{amssymb}
\usepackage[pdfusetitle, bookmarks = true, 
            bookmarksnumbered = false, 
            bookmarksopen, bookmarksopenlevel = 1,
            breaklinks = false, backref = false,
            colorlinks = true,
            linkcolor = blue, urlcolor = blue, citecolor = blue]
            {hyperref}
\usepackage{bm}
\usepackage{upgreek}
\usepackage{braket}
\usepackage[mode = buildnew, subpreambles=false]{standalone}
\date{\today}
\makeatletter
\renewcommand\frontmatter@abstractwidth{\dimexpr\textwidth-10mm\relax}
\makeatother

\begin{document}
\title{Echo-enhanced molecular orientation at high temperatures} 

\author{Ilia Tutunnikov}
\thanks{These authors contributed equally to this work}
\affiliation{AMOS and Department of Chemical and Biological Physics, The Weizmann Institute of Science, Rehovot 7610001, Israel}
\author{Long Xu} 
\thanks{These authors contributed equally to this work}
\affiliation{AMOS and Department of Chemical and Biological Physics, The Weizmann Institute of Science, Rehovot 7610001, Israel}
\author{Yehiam Prior}
\email{yehiam.prior@weizmann.ac.il}
\affiliation{AMOS and Department of Chemical and Biological Physics, The Weizmann Institute of Science, Rehovot 7610001, Israel}
\author{Ilya Sh. Averbukh}
\email{ilya.averbukh@weizmann.ac.il}
\affiliation{AMOS and Department of Chemical and Biological Physics, The Weizmann Institute of Science, Rehovot 7610001, Israel}

\begin{abstract}
    Ultrashort laser pulses are widely used for transient field-free molecular orientation -- a phenomenon important in chemical reaction dynamics, ultrafast molecular imaging, high harmonics generation, and attosecond science. However, significant molecular orientation usually requires rotationally cold molecules, like in rarified molecular beams, because chaotic thermal motion is detrimental to the orientation process. Here we propose to use the mechanism of the echo phenomenon previously observed in hadron accelerators, free-electron lasers, and laser-excited molecules to overcome the destructive thermal effects and achieve efficient field-free molecular orientation at high temperatures. In our scheme, a linearly polarized short laser pulse transforms a broad thermal distribution in the molecular rotational phase space into many separated narrow filaments due to the nonlinear phase mixing during the post-pulse free evolution. Molecular subgroups belonging to individual filaments have much-reduced dispersion of angular velocities. They are rotationally cold, and a subsequent moderate terahertz (THz) pulse can easily orient them. The overall enhanced orientation of the molecular gas is achieved with some delay, in the course of the echo process combining the contributions of different filaments. Our results demonstrate that the echo-enhanced orientation is an order of magnitude higher than that of the THz pulse alone. The mechanism is robust -- it applies to different types of molecules, and the degree of orientation is relatively insensitive to the temperature. The laser and THz pulses used in the scheme are readily available, allowing quick experimental demonstration and testing in various applications. Breaking the phase space to individual filaments to overcome hindering thermal conditions may find a wide range of applications beyond molecular orientation.
\end{abstract}

\maketitle

\section{Introduction \label{sec:Introduction}}

Control over the absolute orientation of molecules in space is essential for a wide variety of physical and chemical processes such as ultra-fast structural imaging, photon-induced processes, or chemical reactions \cite{StapelfeldtSeidman2003,Ohshima2010,Fleischer2012,Lemeshko2013,Koch2019Quantum,Lin2020Review}.
Diverse optical methods have been developed to align and orient molecules in the gas phase under field-free conditions. A well-established route for orienting polar molecules is combining intense non-resonant laser fields with weak electrostatic fields \cite{Friedrich1999,Friedrich1999-2,Sakai2003,Goban2008,Ghafur2009,Holmegaard2009,Mun2014,Takei2016,Omiste2016,Thesing2017}; another one involves using intense two-color laser pulses to induce molecular orientation via interaction with the molecular hyperpolarizability \cite{Vrakking1997,Dion1999,Kanai2001,Takemoto2008,De2009,Oda2010,JW2010,Zhang2011-multicolor,Frumker2012,Spanner2012,Znakovskaya2014,Lin2018All,MelladoAlcedo2020,Shuo2020,xu2021longlasting}.
For chiral molecules, it is possible to induce orientation by using laser pulses with twisted polarization \cite{Yachmenev2016,Gershnabel2018,Tutunnikov2018,Milner2019Controlled,Tutunnikov2019Laser,Tutunnikov2020Observation,Long2022}. With the advance in terahertz (THz) technology, it has become possible to orient polar molecules using single- or half-cycle THz pulses through coupling to the permanent molecular dipole moment \cite{Harde1991,Machholm2001,Fleischer2011,Kitano2013,Egodapitiya2014,Damari2016,Babilotte2016Observation,Babilotte2017,Xu2020,TutunnikovXu2020,Beer2022}.

The efficiency of the above-mentioned methodologies deteriorates fast with increasing rotational temperature, and the molecular orientation is typically just a few percent at room temperature -- too low for many practical applications.  High temperature is detrimental to efficient orientation due to the increased dispersion of the molecular angular velocities with temperature.
In laser-based methods, the degree of orientation may be enhanced by simply increasing the field intensity. Still, this approach has a practical limit because high-intensity laser pulses, especially two-color fields, give rise to significant molecular ionization, an undesired consequence of the strong field. On the other hand, in the case of THz pulses, the available field strengths are insufficient for achieving a sizable degree of orientation. Thus, attaining efficient molecular orientation at room temperatures is still an open, very challenging problem. 

Over the years, several sophisticated proposals for enhancing molecular alignment and orientation have been put forward. These include applying optimized sequences of pulses \cite{Averbukh2001,Averbukh2003,Averbukh2004,Zhang2011Field,Zhang2016},
optimizing pulses' parameters \cite{Nakajima2012Optimal,Mun2018,Mun2019,Mun2020,Maruf2020,Yu2021,Maruf2022,Wang2020,Zhou2021,Long2022optimal}, and combining one-color laser pulses with two-color and THz pulses
\cite{Spanner2004,Daems2005,Gershnabel2006,Gershnabel2006-2,Kitano2011,Zhang2011Controlling,Tehini2012Field,Shu2013Field,Liu2013,Tao2018}. Several schemes were experimentally implemented \cite{Bisgaard2004,Pinkham2007,Egodapitiya2014,Sonoda2018} and demonstrated improved orientation, mainly in the low-temperature regime.

Here, we propose an efficient and general approach for achieving significant field-free molecular orientation by moderate THz pulses even at room temperature. The method uses a pair of time-delayed pulses arranged in the echo configuration. The first, non-resonant laser pulse transforms a broad and smooth thermal probability-density distribution in the molecular rotational phase-space into multiple filaments (narrow strips). The second (typically weak) delayed THz excitation is then applied to the ``filamented'' molecular ensemble and causes a significant orientation of molecular groups belonging to different narrow filaments, each evolving independently. The enhanced ensemble-averaged orientation is achieved after some additional delay when the contributions of the independent filaments come together by the mechanism of echo formation. We show that for a proper choice of the excitation parameters, the amplitude of the orientation echo may be an order of magnitude higher than the orientation produced by the THz pulse alone. Moreover, the ``filamenting'' prepulse suppresses the detrimental effect of the molecular thermal rotational motion, making the echo amplitude rather insensitive to elevated temperatures.

Echoes are known in various domains of physics. The most notable
example is Hahn's spin-echo \cite{Hahn1950PR,Hahn1953} induced by
two delayed magnetic field pulses. They result in a delayed magnetization response appearing after an additional delay equals the delay between the pulses. Various types of echoes have been discovered and studied in other fields, including photon echoes \cite{Kurnit1964,Mukamel1995}, neutron spin-echo \cite{Mezei1972}, cyclotron echoes \cite{Hill1965}, plasma-wave
echoes \cite{Gould1967}, echoes in optically trapped cold atoms \cite{Bulatov1998,Buchkremer2000,Herrera2012}, echoes in particle accelerators \cite{Stupakov1992,Spentzouris1996,Stupakov2013,Sen2018},
and echoes in free-electron lasers \cite{Hemsing2014}. In addition,
echoes in single quantum systems have been discussed \cite{Kerr2021,Bouncer2021} and observed \cite{Morigi2002,Meunier2005,Qiang2020}.

The enhancement mechanism of the present paper is related to the recently
predicted and observed molecular alignment echoes \cite{Karras2015,Karras2016,Lin2016,Lin2017,Rosenberg2018,Zhang2019,Ma2019,Hartmann2020,Ma2020Ultrafast,Lin2020Review,Xu2020UDRecho,Ma2021}.
In contrast to the spin echoes \cite{Hahn1950PR,Hahn1953}, they are
based on the phenomenon of phase space filamentation known in non-linear
systems, e.g., in the dynamics of stellar systems \cite{Lynden-Bell1967}
and in accelerator physics \cite{Guignard1988,Stupakov1992,Stupakov2013}.
Our paper presents the first application of the filamentation-based
echoes to the problem of molecular orientation.

The paper is organized as follows. We begin with a qualitative physical description
of the scheme using a simplified model of rigid planar rotors. Then
we present the results of three-dimensional classical and fully quantum
simulations of the echo-enhanced orientation of linear and symmetric-top
molecules at room temperature. Finally, we summarize the results and
discuss the prospects of the proposed scheme.

\section{Qualitative discussion \label{sec:Qualitative-discussion}}

To understand the physical principles behind the enhanced orientation, we start with a simplified planar rotor model, where a linear polar molecule is modeled as a rigid rotor restricted to rotate in the $XY$ plane. To begin with, let us first consider the orientation dynamics induced by a single orienting kick. The orienting interaction potential is $\propto-\cos(\theta)$, where $\theta$ is the angle between the molecular axis and the $X$-axis. Such a potential describes, e.g.,
the interaction of a single- or half-cycle THz pulse with the permanent molecular dipole.
\begin{figure}
    \centering
    \includegraphics{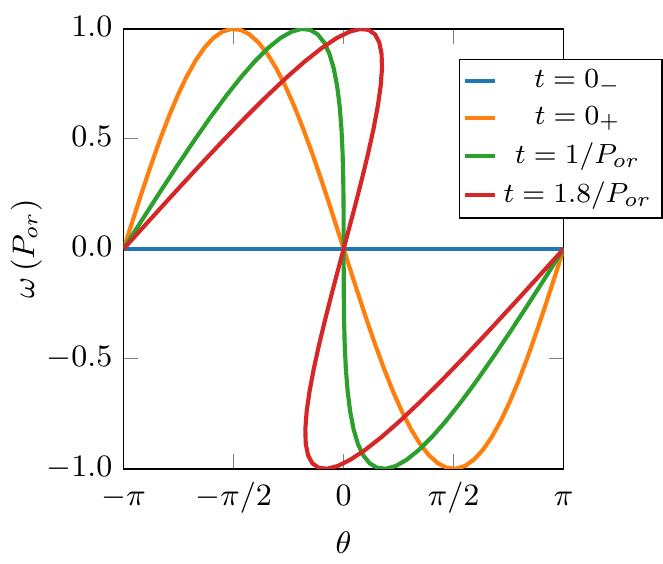}
    \caption
        {
         Phase space following a single orienting kick at zero initial rotational
         temperature. 
         \label{fig:FIG1-zero-T}
         }
\end{figure}
\paragraph{Single orienting kick --}

In the impulsive approximation (when the change in $\theta$ during
the pulse is negligible), $\theta$ and the molecule's angular velocity,
$\omega$ after the orienting kick are given by 
\begin{equation}
    \theta(t) = \theta_{0}+\omega t;\quad\omega=\omega_{0}-P_{or}\sin(\theta_{0}),
    \label{eq:theta-omega-ori}
\end{equation}
where $\theta_{0}$ and $\omega_{0}$ are the initial angle and angular
velocity. $P_{or}$ is proportional to the amplitude of the orienting
field, $\varepsilon_{0,or}$ [for details, see Appendix \ref{sec:App-A-2D-model-details},
Eq. \eqref{eq:App-A-P-THz}] and has a physical meaning of the molecular angular velocity change due to the kick.

To study the rotational dynamics of an ideal molecular gas, we consider
a collection of planar rotors that are initially isotropically distributed
in the $XY$ plane. At zero initial temperature, the phase space density
before the kick is presented by a horizontal line $\omega=0$ (see
Fig. \ref{fig:FIG1-zero-T}, $t=0_{-}$). The orienting pulse transforms
the initial distribution into a sine-shaped line ($t=0_{+}$). With
time, the curve folds as particles with opposite angular velocities
move in opposite directions along the $\theta$ axis. The ensemble-averaged
degree of orientation, or the orientation factor $\braket{\cos(\theta)}$
reaches a universal maximum $\approx58\%$, independent of $P_{or}$
at $t=1.8/P_{or}$.

\begin{figure}
    \includegraphics{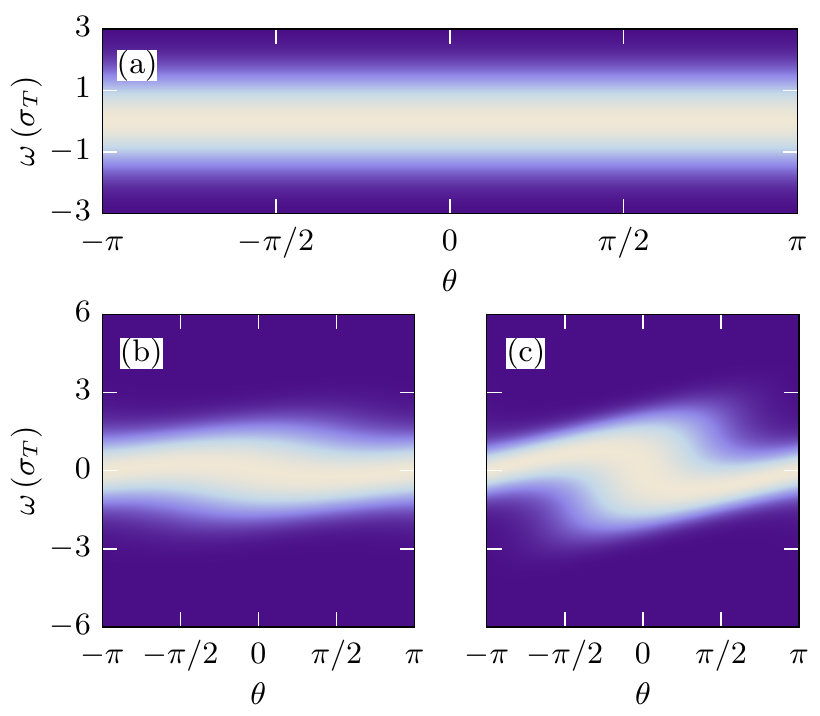}
    \caption
        {
        Phase space distribution before and after  a single orienting pulse. (a)
        At  $t=0$, before the kick. (b) After the pulse at $t=\sigma_{T}^{-1}$ for a relatively weak orienting pulse $P_{or}=0.25\sigma_{T}$. (c) After the pulse at $t=\sigma_{T}^{-1}$ for a stronger orienting pulse $P_{or}=0.75\sigma_{T}$.
        \label{fig:FIG2-phase-space-THz-short}
        }
\end{figure}

At finite, non-zero initial rotational temperature, the initial phase space density is given by $p(\theta,\omega,t=0)=(2\pi)^{-3/2}\sigma_{T}^{-1}\exp[-\omega_{0}^{2}/(2\sigma_{T}^{2})]$,
see Fig. \ref{fig:FIG2-phase-space-THz-short}(a). The parameter $\sigma_{T}=\sqrt{k_{B}T/I}$ defines the thermal dispersion of angular velocities, where $T$ is the temperature, $k_{B}$ is the Boltzmann constant and $I$ is the moment of inertia of a single molecule.
In the case of the initial thermal ensemble, the maximum degree of orientation is determined by the ratio $P_{or}/\sigma_{T}\propto\varepsilon_{0,or}/\sqrt{T}$. When the orienting kick is relatively 
strong (or, equivalently, the temperature is relatively low), the resulting orientation is high, and vice versa, when the kick strength is insufficient to overcome the thermal dispersion of angular velocities.
The orientation factor after the kick is given by [for derivation, see Appendix
\ref{sec:App-B-single-ori}, Eq. \eqref{eq:App-B-ori-factor-THz}]
\begin{equation}
    \braket{\cos(\theta)}(t) = e^{-\sigma_{T}^{2}t^{2}/2}J_{1}(P_{or}t),
    \label{eq:ori-factor-THz}
\end{equation}
where $J_{1}(z)$ is the Bessel function of order one. When time is
measured in units of $1/\sigma_{T}$, the orientation factor depends
on the parameter $P_{or}/\sigma_{T}$.
Unfortunately, the currently available THz pulses result in a weak deformation of the phase space density at room temperature. Shortly after a typical orienting kick, the phase space density is delocalized and looks like in Fig. \ref{fig:FIG2-phase-space-THz-short}(b). When the orienting kick is relatively strong (or, equivalently, the temperature is relatively low), the phase space distribution folds like in Fig. \ref{fig:FIG2-phase-space-THz-short}(c). 

\paragraph{Phase space filamentation --}
To help a weak orienting pulse overcome the detrimental effect of high temperatures, we suggest stratifying the rotational phase space density into numerous separated strips (filaments). This way, the orienting pulse interacts with many individual narrow filaments rather than a single broadly dispersed thermal distribution. For stratification, we use a relatively strong prepulse -- a nonresonant linearly polarized short laser pulse that precedes the orienting one. Intense laser pulses interacting with molecular polarizability have been widely used for inducing transient field-free molecular alignment \cite{StapelfeldtSeidman2003,Ohshima2010,Fleischer2012,Lemeshko2013,Koch2019Quantum}.
However, here we are not interested in the alignment effect but focus on the post-alignment stage characterized by highly stratified rotational phase space density \cite{Karras2015,Lin2016,Lin2017} (see below).

\begin{figure}
    \centering
    \includegraphics{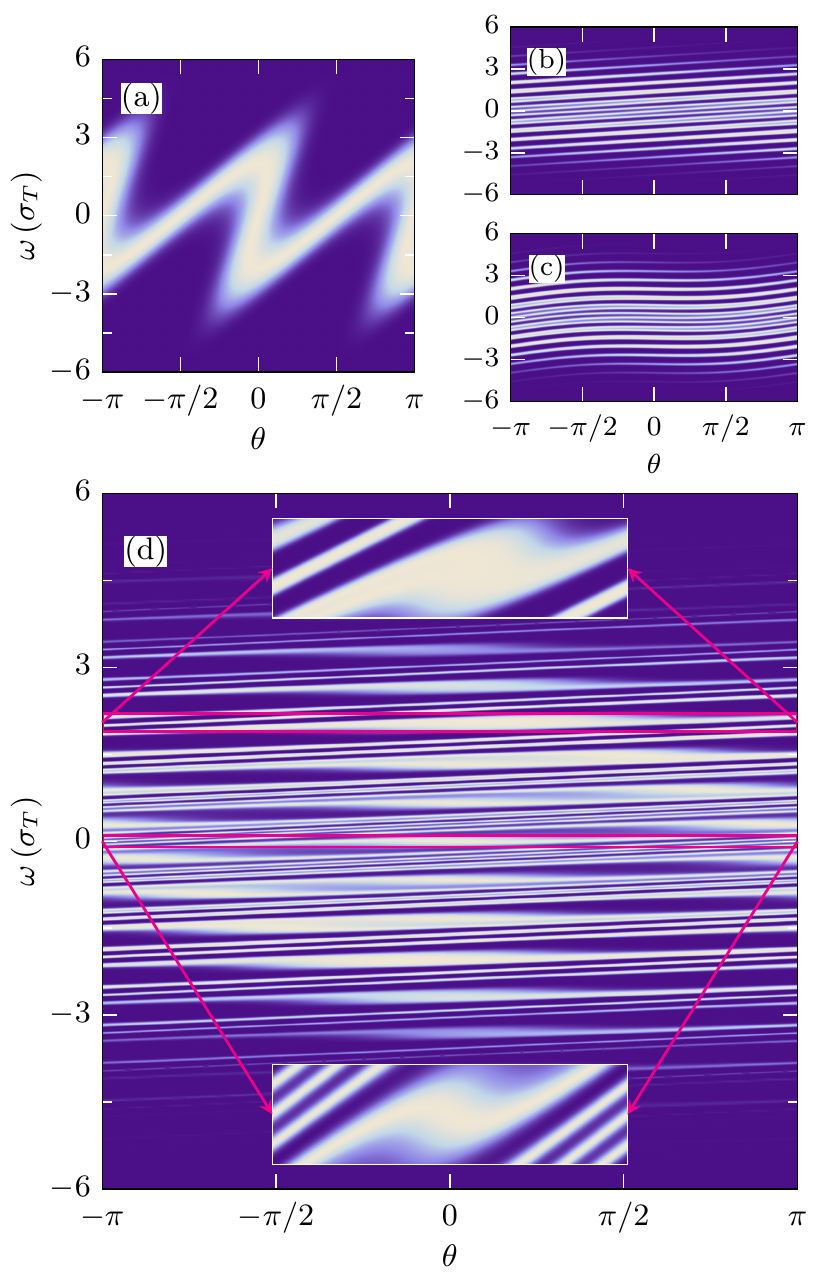}
    \caption
        {
        Phase space distribution. 
        (a) Shortly after the stratifying kick, $t=0.4/\sigma_{T}$. 
        (b) Just before the orienting kick, $t=\tau_{-}$, for $\tau=5/\sigma_{T}$. 
        (c) Just after the orienting kick, $t=\tau_{+}$. 
        (d) Orientation along the $X$ axis, before the time of the orientation echo at $t=3\tau$. 
        The regions marked in magenta are shown in the insets.  
        Here $P_{pre}=2\sigma_{T}$, $P_{or}=0.25\sigma_{T}$.
        \label{fig:FIG3-phase-space-al-or}
        }
\end{figure}

A prepulse applied at $t=0$ transforms the rotors variables according to
\begin{equation}
    \theta(t) = \theta_0 + \omega t;
    \quad
    \omega = \omega_0 - P_{pre}\sin(2\theta_0),
    \label{eq:theta-omega-ali}
\end{equation}
where $P_{pre}$ is proportional to the intensity of the laser prepulse
[for details, see Appendix \ref{sec:App-A-2D-model-details}, Eq. \eqref{eq:App-A-P-laser}]. 
The phase space density after the prepulse is described by 
\begin{equation}
    \hspace{-2mm}
    p(\theta,\omega,t)=\frac{\sigma_{T}^{-1}}{(2\pi)^{3/2}}
    \exp\left[-\frac{(\omega+P_{pre}\sin[2\theta-2\omega t])^2}{2\sigma_T^2}\right]\!,\!\!
    \label{eq:phase-space-density-al}
\end{equation}
where we used Eq. \eqref{eq:theta-omega-ali} to express $\omega_{0}$
in terms of $\omega$ and $\theta$ after the pulse, and substituted
$\omega_{0}$ into the Boltzmann distribution. The stratifying pulse
imprints a sinusoidal shape on the phase-space density which folds with time. Shortly after the excitation, the phase-space distribution
concentrates near $\theta=0,\pi$, corresponding to alignment along
the $X$ axis [see Fig. \ref{fig:FIG3-phase-space-al-or}(a)].
It's important for our discussion that with time, the phase space density undergoes filamentation and turns into a series of almost parallel filaments (strips) separated  in angular velocity by $\pi/t$ [see Eq. \eqref{eq:phase-space-density-al}]. Note that longer waiting times result in narrower filaments because the phase space volume is conserved.

After a delay $\tau$ (at $t=\tau$), when the filaments are numerous
and thin, [see Fig. \ref{fig:FIG3-phase-space-al-or}(b)], we apply the orienting THz kick. Now, each filament resembles the phase-space density of an isotropic cold molecular ensemble (see Fig. \ref{fig:FIG1-zero-T}, $t=0_{-}$). Moreover, since the dispersion of angular velocities within each filament is much smaller than the initial thermal dispersion, the interaction with the weak orienting pulse results in non-negligible deformation of the filaments [see Fig. \ref{fig:FIG3-phase-space-al-or}(c)]. 

With time, each filament folds like in Fig. \ref{fig:FIG2-phase-space-THz-short}(c)
developing a bunch. This folding does not result in substantial orientation just after the orienting kick because the bunches move relative to each other (they are spaced apart by $\pi/\tau$ along the $\omega$ axis).
Due to the ``quasi-quantization'' of the angular velocities, the
folded filaments synchronize near $\theta=0,\pi$ after an additional
time delay equal to $\tau$ (at $t\approx2\tau$, not shown). 
This synchronization manifests itself in alignment echo \cite{Karras2015,Karras2016,Lin2016,Lin2017,Lin2020Review,Xu2020UDRecho}. Similar synchronization dynamics is behind the echo phenomenon in several other nonlinear systems \cite{Stupakov1992,Stupakov2009,Stupakov2013,Qiang2020,Kerr2021,Bouncer2021}.

After an additional delay of $\tau$ (at $t\approx3\tau$), we witness
another echo, and this is a new observation. The individual bunches re-synchronize in an asymmetric manner resulting in molecular orientation. First, the bunches accumulate
near $\theta=0$ [just before $t=3\tau$, see Fig. \ref{fig:FIG3-phase-space-al-or}(d)], and then near $\theta=\pi$ [just after $t=3\tau$, not shown]. The closeups in Fig. \ref{fig:FIG3-phase-space-al-or}(d) demonstrate that the bunches in each filament look similar to those seen in Fig.
\ref{fig:FIG2-phase-space-THz-short}(c). Such re-synchronizations happen again at later times: $t=5\tau,\,7\tau,\dots$, and manifest themselves in orientation echoes of higher orders.

\begin{figure}
    \centering
    \includegraphics{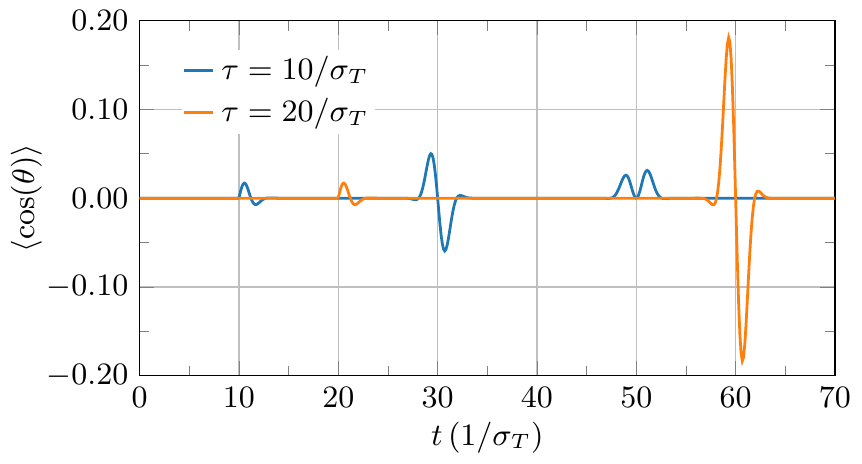}
    \caption
        {
        Orientation factor following combined excitation by stratifying and orienting pulses. 
        Here, $P_{pre}=2\sigma_{T}$, $P_{or}=0.1\sigma_{T}$. 
        Note the enhanced orientation at $t=2\tau$ for both cases of $\tau=10/\sigma_{T}$ 
        and $\tau=20/\sigma_{T}$, with the higher degree of orientation for the longer delay.
        With the orienting pulse alone, the maximal orientation factor is $\approx0.03$.
        \label{fig:FIG4-ori-factors}
        }
\end{figure}

Figure \ref{fig:FIG4-ori-factors} shows the time-dependent orientation factor, $\braket{\cos(\theta)}$ for two delays, $\tau=10/\sigma_{T}$ and $\tau=20/\sigma_{T}$. In both cases, the first orientation echo emerges after a delay equal to $2\tau$. However, compared to $\tau=10/\sigma_{T}$, the echo amplitude for  $\tau=20/\sigma_{T}$ is significantly higher. In line with the qualitative arguments presented above, longer delays give rise to  thinner filaments which, in turn, ``helps'' the weak
orienting kick.

The orientation factor in the presence of both stratifying and orienting
kicks is given by [for derivation, see Appendix \ref{sec:App-B-both-pre-ori}, Eq. \eqref{eq:App-B-ori-after-lsr-THz}]
\begin{equation}
    \braket{\cos(\theta)}(t) = \Theta(t-\tau)\sum_{\mathrm{odd}\,k}^{\infty}S_k(t),
    \label{eq:ori-after-al-or}
\end{equation}
where $\Theta(t-\tau)$ is the step-function and 
\begin{equation}
    \!S_k(t)=e^{-\sigma_T^2(t-k\tau)^2/2}J_{k}[P_{or}(t\!-\!\tau)]J_{\frac{k-1}{2}}[P_{pre}(k\tau\!-\!t)].
\end{equation}
The orientation signal consists of a series of pulsed responses separated
in time by $2\tau$. The first orientation echo $(k=3)$ is proportional
to
\begin{equation}
    \!\!\!S_{3}(t)\approx e^{-\sigma_T^2(t-3\tau)^2/2}J_3(2P_{or}\tau)J_1[P_{pre}(3\tau-t)].
\end{equation}
Here we used the facts that $J_{3}(z)$ increases with $z$ for $z<4$,
and $t-3\tau$ is relatively small (due to the Gaussian factor). This
expression shows the essential feature of the proposed scheme --
for fixed temperature and stratifying kick (prepulse) strength, $\sigma_{T}$
and $P_{pre}$, the echo amplitude increases with $P_{or}\tau$. In
other words, increasing the delay $\tau$ is equivalent to increasing
the orienting kick strength.

\begin{figure}
    \centering
    \includegraphics{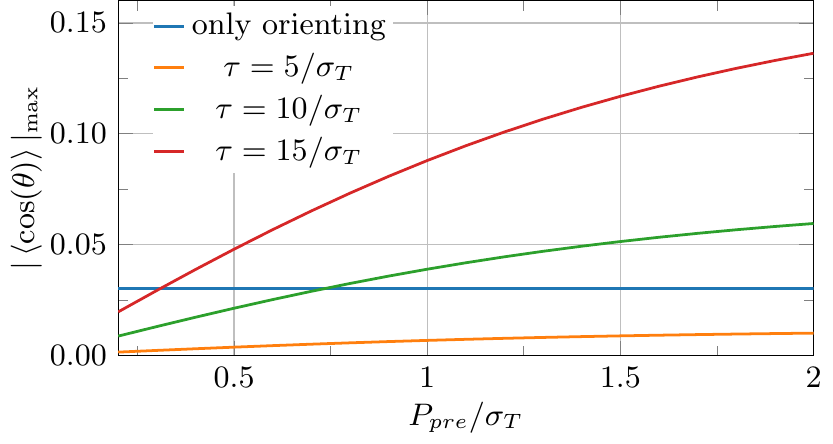}
    \caption
        {
        Maximum orientation factor (absolute value) as a function of $P_{pre}$ for fixed $P_{or}=0.1/\sigma_{T}$. 
        Note that for the low intensity prepulses, the degree of orientation is lower than 
        in the absence of a prepulse altogether. This is a result of  very minimal filamentation 
        caused by the weak prepulses, which, in effect, slightly increases the dispersion 
        of rotational velocities.
        \label{fig:FIG5-max-ori-afo-Plsr}
        }
\end{figure}

Moreover, the echo amplitude increases with $P_{pre}$ as well. Figure \ref{fig:FIG5-max-ori-afo-Plsr} shows the maximum orientation factor for several delays as a function of $P_{pre}$ (for fixed $P_{or}$). When compared to the maximal orientation achieved by the orienting kick alone, the addition of the prepulse enhances the orientation starting from certain $P_{pre}$ values (depending on the delay, $\tau$). Note that if the prepulse is weak, it has a negative effect, namely the degree of orientation is even lower than when using only the orienting pulse. The reason is that a weak prepulse does not generate narrow well-separated filaments which can be distorted by the weak orienting pulse. Instead, it effectively slightly increases the dispersion of rotational velocities.

Terms with $k>3$ in Eq. \eqref{eq:ori-after-al-or} correspond to the orientation echoes of a higher order that typically (but not always) have lower amplitude compared to the first echo \cite{Stupakov2009,Karras2015,Karras2016,Lin2016,Lin2017,Lin2020Review}.

In the classical case, the delay $\tau$ could, in principle, be increased indefinitely to achieve higher orientation. However, when the delay becomes comparable to the time of rotational quantum revivals \cite{Averbukh1989,Robinett2004}, the interplay between echoes and
revivals needs to be taken into account. As was shown previously \cite{Herrera2012,Kerr2021}, a particular type of echo exists, termed \textit{quantum echo}. A series of quantum echoes emerge on a long time scale just before the quantum revivals. These echoes have a valuable property -- in the limit of weak echo-inducing excitation (the orienting kick in our case), their amplitude is higher than the classical echoes discussed. We take advantage of this to achieve even higher molecular orientation with constrained delays.

\begin{figure}
    \centering
    \includegraphics{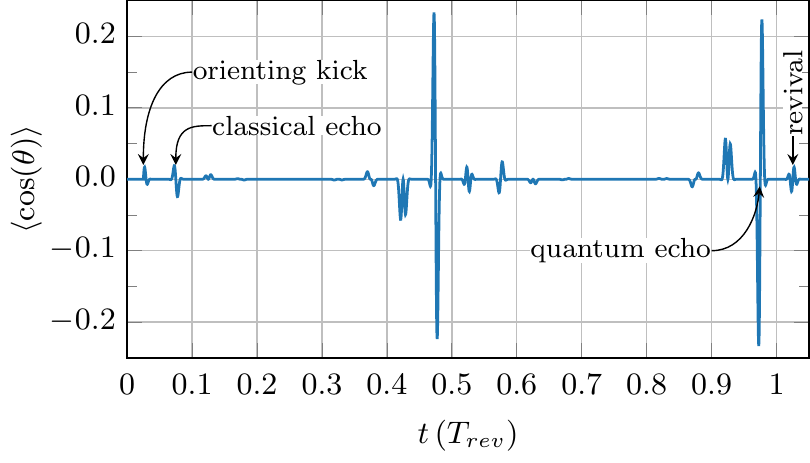}
    \caption
        {
        Orientation factor calculated quantum mechanically at $T=300\,\mathrm{K}$. 
        Here, the orienting kick is applied after a delay $\tau=0.025T_{rev}$, 
        which corresponds to $\tau\approx7.2/\sigma_{T}$. The excitation strengths, 
        $P_{pre}$ and $P_{or}$ are the same as in Fig. \ref{fig:FIG4-ori-factors}. 
        When only the orienting kick is applied, the maximal orientation is $\approx0.03$.
        \label{fig:FIG6-quantum-ori}
        }
\end{figure}

To illustrate this point, we consider the dynamics of a kicked quantum rigid 2D rotor (for details, see Appendix \ref{sec:App-C-Quantum-rigid-rotor}). The rotational revival time of the quantum planar rotor is $T_{rev}=4\pi I/\hbar$, where $I$ is the moment of inertia. Figure \ref{fig:FIG6-quantum-ori} shows the orientation factor in the quantum model in the presence of the stratifying (prepulse) and orienting kicks. Here, for demonstration purposes, we choose a short time delay $\tau=0.025T_{rev}$ resulting in a relatively weak classical orientation echo at $t=3\tau=0.075T_{rev}$. The amplitude of the classical echo is comparable to the maximal orientation obtained when using the orienting pulse alone, $\approx3\%$. In striking contrast, a quantum echo emerges at a delay $-2\tau$ before
the revival, at $t=0.975T_{rev}$ with a much higher amplitude of $\approx22\%$. Notice that echoes of the same magnitude (but opposite sign) emerge just before the half revival.
Alignment echoes emerging before the quantum revivals, the so-called imaginary echoes, were studied by us both theoretically and experimentally \cite{Lin2016,Lin2017}.

\section{Three-dimensional simulations \label{sec:Quantum-simulations}}

\subsection{Numerical methods \label{sec:Numerical methods}}

\begin{figure}
    \centering
    \includegraphics{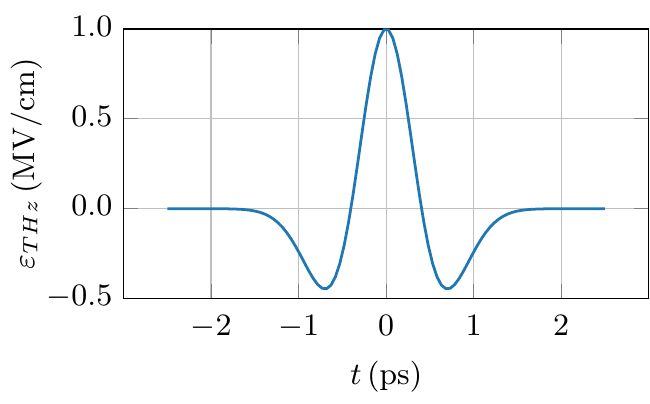}
    \caption
        {
        Electric field of the THz pulse, given by 
        $\varepsilon_{THz}(t)=\varepsilon_{0,THz}(1-2\kappa t^{2})\exp(-\kappa t^{2})$.
        Here, $\varepsilon_{0,THz}=1.0\,\mathrm{MV/cm}$, $\kappa=3.06\,\mathrm{ps}^{-2}$.
        \label{fig:FIG7-THz-field}
        }
\end{figure}

So far, we considered the orientation enhancement in the simplified
model of 2D rigid rotors. In this section, we apply the developed understanding to ``real world'' by considering the corresponding
three-dimensional classical and quantum dynamics of linear and prolate symmetric-top molecules driven by the experimentally available laser and THz fields. The rotational dynamics is treated within the
rigid rotor approximation. We model the time-dependent electric field, consisting of the delayed non-resonant laser and single-cycle THz pulses, using 
\begin{equation}
    \mathbf{E}(t)=\varepsilon_{lsr}(t)\cos(\omega t)\bm{e}_{Z}+\varepsilon_{THz}(t-\tau)\bm{e}_{Z},
    \label{eq:electric-field}
\end{equation}
where $\omega$ is the carrier frequency of the laser pulse, $\varepsilon_{lsr}(t)=\varepsilon_{0,lsr}\exp[-2\ln2(t/\sigma_{lsr})^{2}]$, $\varepsilon_{0,lsr}$ is the peak amplitude, and $\sigma_{lsr}$ is the full width at half maximum (FWHM) of the laser pulse intensity profile. $\varepsilon_{THz}(t)=\varepsilon_{0,THz}(1-2\kappa t^{2})\exp(-\kappa t^{2})$ \cite{Babilotte2016Observation,Babilotte2017}, where $\varepsilon_{0,THz}$ is the peak amplitude of the THz pulse, $\kappa$ determines the temporal width of the THz pulse. $\tau$ is the time delay between the laser and THz pulses and $\mathbf{e}_{Z}$ is a unit vector along the laboratory $Z$ axis. Figure \ref{fig:FIG7-THz-field} shows the electric field
amplitude of the THz pulse as a function of time.
\paragraph{Classical simulation --}
\noindent We use the Monte Carlo method to simulate the behavior of
a classical ensemble. Initially, $N=10^{7}$ sample molecules are
isotropically distributed in space. Their angular velocities are given
by the Boltzmann distribution $P(\Omega_{i})\propto\exp[-I_{i}\Omega_{i}^{2}/(2k_{B}T)]$,
where $i=x,y,z$ refers to the molecular principal axes of inertia.
In the case of a linear molecule $I_{z}=0$, $I_{x}=I_{y}\equiv I$,
while in the case of a prolate symmetric-top molecule, $I_{z}<I_{x}=I_{y}\equiv I$.
For each sample molecule, the rotational dynamics in the rotating
molecular frame is described by Euler's equations \cite{Goldstein2002Classical}
\begin{equation}
    \mathbf{I}\bm{\dot{\Omega}}=(\mathbf{I}\bm{\Omega})\times\bm{\Omega}+\mathbf{T},
    \label{eq:Eulers-equations}
\end{equation}
where $\mathbf{I}=\mathrm{diag}(I_{x},I_{y},I_{z})$ is the moment
of inertia tensor, $\bm{\Omega}=(\Omega_{x},\Omega_{y},\Omega_{z})$
is the angular velocity, and $\mathbf{T}=(T_{x},T_{y},T_{z})$ is
the torque due to the interactions with the external electric field.
The laser field interacts with the molecular polarizability $\bm{\alpha}=\mathrm{diag}(\alpha_{xx},\alpha_{yy},\alpha_{zz})$,
and the torque is given by $\ensuremath{\mathbf{T}=[\bm{\alpha}\mathbf{E}_{mol}(t)]\times\mathbf{E}_{mol}(t)}$.
The THz field interacts with the molecular permanent dipole moment,
$\boldsymbol{\mu}$, and the torque is given by $\ensuremath{\mathbf{T}=\boldsymbol{\mu}\times\mathbf{E}_{mol}(t)}$.
Here, $\mathbf{E}_{mol}(t)$ is the representation of the electric
field vector in the basis of the molecular principal axes. The relation
between the laboratory and the rotating molecular frames is parametrized
by a quaternion, $q=(q_{0},q_{1},q_{2},q_{3})$ \cite{Kuipers1999Quaternions,Coutsias2004The}.
The quaternion's equation of motion is $\dot{q}=q\Omega/2$, where
the quaternions multiplication rule is assumed, and $\Omega=(0,\Omega_{x},\Omega_{y},\Omega_{z})$.
The orientation factor at time $t$ is obtained by averaging over
all sample molecules (ensemble average). A more detailed description
of the classical simulations can be found in \cite{Tutunnikov2019Laser,xu2021longlasting}.
\paragraph{Quantum simulation --}
\noindent The Hamiltonian describing the rotational dynamics in the
presence of external fields {[}see Eq. \eqref{eq:electric-field}{]}
is \cite{Krems2018Molecules,Koch2019Quantum} 
\begin{align}
    H(t) & =H_{r}-\mu\varepsilon_{THz}(t)\cos(\theta)-\frac{\Delta\alpha}{4}\varepsilon_{lsr}^{2}(t)\cos^{2}(\theta),
\end{align}
where $H_{r}$ is the rotational kinetic energy Hamiltonian, $\mu=|\boldsymbol{\upmu}|$
is the permanent molecular dipole moment, $\Delta\alpha=\alpha_{zz}-\alpha_{xx}=\alpha_{zz}-\alpha_{yy}$
is the polarizability anisotropy, and $\theta$ is the polar angle between
the molecular symmetry axis and the laboratory $Z$ axis (field polarization
axis). The polarizability interaction term is averaged over the optical
cycle.

In the simulations, the wave function $\Psi$ is expressed in the
basis of the eigenfunctions of $H_{r}$. For a linear molecule, these
are the spherical harmonics $Y_{J}^{M}(\theta,\phi)$ ($\phi$ is
the azimuthal angle), while for a symmetric-top molecule, these are
Wigner D-functions $D_{MK}^{J}(\theta,\phi,\chi)$, where $\text{\ensuremath{\chi}}$
is the angle of rotation about the molecular symmetry axis. The quantum
numbers $J$, $M$ and $K$ correspond to the magnitude of the angular
momentum, projection of angular momentum on the laboratory $Z$ axis,
and projection of the angular momentum on the molecular symmetry axis.
The eigenenergies are $E_{J}=BJ(J+1)$ (linear molecule) and $E_{JK}=CJ(J+1)+(A-C)K^{2}$
(symmetric-top molecule), where $B=C=\hbar^{2}/(2I)$ and $A=\hbar^{2}/(2I_{z})$.
The time-dependent Schr{\"o}dinger equation is solved by numerical
exponentiation of the Hamiltonian matrix (see Expokit \cite{sidje1998Expokit}).

In the case of the linear molecule, the orientation factor is given
by the averaged expectation value 
\begin{align}
    \braket{\cos(\theta)}(t)=\frac{1}{Z}\sum\limits _{JM}\braket{\cos(\theta)}_{JM}(t)e^{-E_J/(k_B T)},
    \label{eq:Quantum-Boltzmann-distribution}
\end{align}
where $Z$ is the partition sum, $\braket{\cos(\theta)}_{JM}(t)=\braket{\Psi_{JM}(t)|\cos(\theta)|\Psi_{JM}(t)}$
for the initial state $\Psi_{JM}(t=0)=Y_{J}^{M}$, and the sum runs
over all the thermally populated eigenstates $Y_{J}^{M}$.

In the case of symmetric-top molecules, due to the increased number
of thermally populated states, we use the random phase wave functions
method (see, e.g. \cite{Kallush2015}) to simulate the behavior of
the thermal ensemble. The orientation factor is given by the average
\begin{align}
    \braket{\cos(\theta)}(t)=\frac{1}{N}\sum\limits _{n=1}^{N}\braket{\cos(\theta)}_{n}(t),
    \label{eq:Quantum-Boltzmann-distribution-1}
\end{align}
where $N$ is the number of initial states used, $\braket{\cos(\theta)}_{n}(t)$
is the orientation factor obtained for the initial state $\psi_{n}$
\begin{align}
\psi_{n}=\sum_{JKM}\sqrt{\frac{\epsilon_{K}e^{-E_{JK}/(k_{B}T)}}{Z}}D_{MK}^{J}e^{i\varphi_{n,JKM}},
\end{align}
where $\varphi_{n,JKM}\in[0,2\pi)$ is a random number and the sum
runs over all the thermally populated eigenstates $D_{MK}^{J}$. For
$\mathrm{CH_{3}I}$ molecule, the statistical weight due to the nuclear
spin statistics is given by \citep{McDowell1990Rotational} 
\begin{align}
    \epsilon_{K}=\frac{(2I_{H}+1)^{3}}{3}\left[1+\frac{2\cos(2\pi K/3)}{(2I_{H}+1)^{2}}\right],
    \label{eq:statistic-weight}
\end{align}
with $I_{H}=1/2$.
\subsection{Results: linear molecule \label{sec:Results}}
We consider $\mathrm{OCS}$ as an example linear molecule. The molecular
properties were taken from NIST (DFT, method: CAM-B3LYP/aug-cc-pVTZ)
\cite{johnson1999nist}: $I=83.021\,\mathrm{amu\,\mathring{A}^{2}}$,
$\mu=0.755\,\mathrm{Debye}$, and $\Delta\alpha=3.738\,\mathrm{\mathring{A}^{3}}$.
The rotational revival time of the OCS molecule is $T_{rev}=1/(2cB)\approx82.4\,\mathrm{ps}$,
where $c$ is the speed of light.

\begin{figure}
    \centering
    \includegraphics{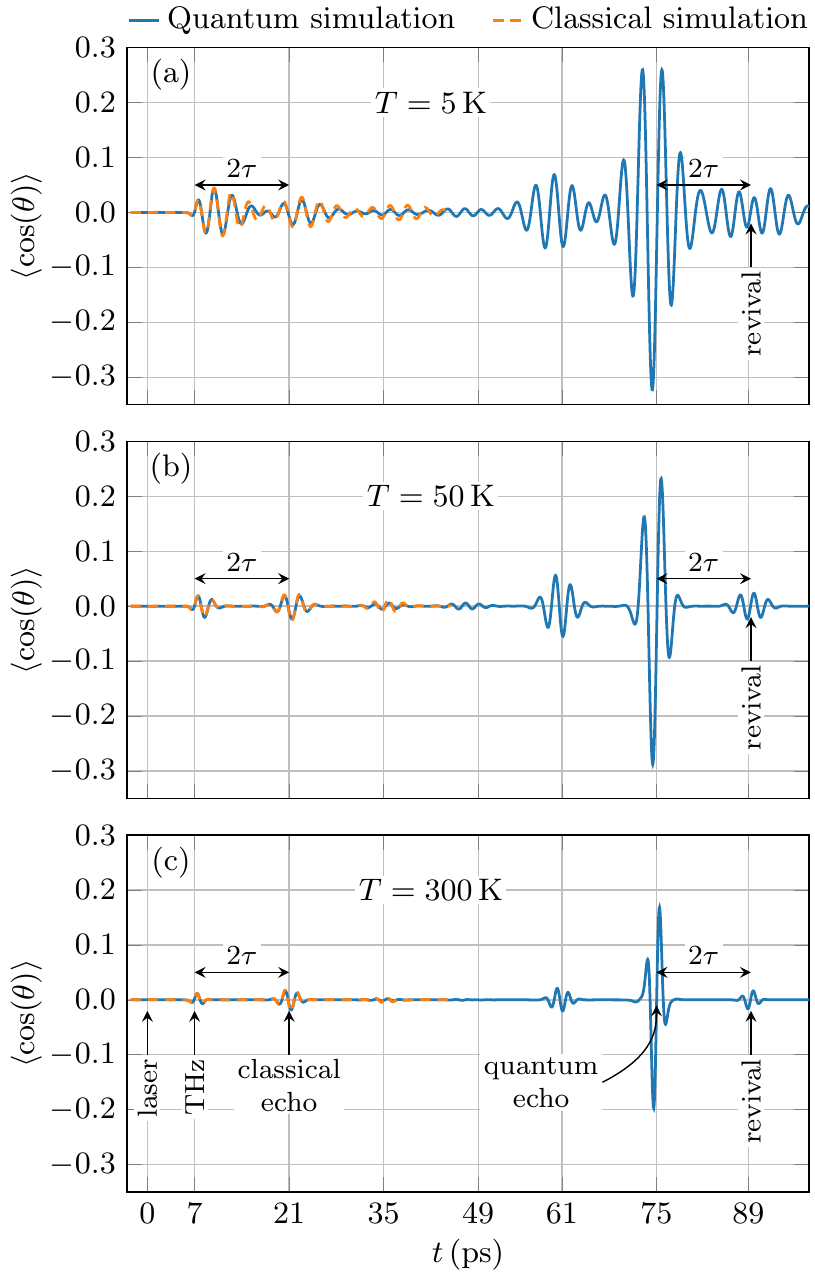} 
    \caption
        {
        Orientation factor for OCS molecules at different temperatures. The molecules undergo combined excitation by stratifying pulse followed by a THz pulse with a delay $\tau=7\,\mathrm{ps}$. The temperatures are: (a) 5\,K, (b) 50\,K, and (c) 300\,K,
        and the maximum orientation factors obtained without the stratifying laser pulses are 0.0074, 0.0201, and 0.0215, respectively.
        \label{fig:FIG8-OCS-temps}
        }
\end{figure}

Figure \ref{fig:FIG8-OCS-temps} shows the time-dependent orientation
factor, $\braket{\cos(\theta)}(t)$ obtained using classical and quantum
simulations. The molecules were excited by the delayed laser and THz
pulses {[}see Eq. \eqref{eq:electric-field}{]} at different initial
rotational temperatures. The delay between the pulses was set to $\tau=7\,\mathrm{ps}$.
The peak intensity of the laser pulse is $I_{0,lsr}=4\times10^{13}\,\mathrm{W/cm^{2}}$,
FWHM is $\sigma_{lsr}=100\,\mathrm{fs}$. The THz pulse parameters
are $\varepsilon_{0,THz}=1.0\,\mathrm{MV/cm}$, $\kappa=3.06\,\mathrm{ps}^{-2}$.
Laser and THz pulses with similar parameters are readily available
nowadays \cite{Clerici2013Wavelength,Oh2014Generation,Vicario2014Infrared,Shalaby2015Demonstration}.

Both quantum and classical simulations demonstrate an immediate orientation
response emerging during and shortly after the THz pulse excitation
(at $t\approx7\,\mathrm{ps}$). The first classical echo appears with
a delay $2\tau$ after the THz pulse, at $t\approx21\,\mathrm{ps}$.
The quantum mechanical result shows a strong echo signal before the
orientation quantum revival, at $t\approx75.4\,\mathrm{ps}$. The
magnitude of this echo is an order of magnitude larger than the maximum
orientation induced by THz pulse alone. In agreement with the previous
studies \cite{Herrera2012,Kerr2021} and the qualitative analysis
based on the simplified models (see Sec. \ref{sec:Qualitative-discussion}),
the orientation during the quantum echo is significantly higher compared
to the orientation during the classical echo. The magnitude of the
quantum echo dominates at all temperatures represented in Fig. \ref{fig:FIG8-OCS-temps},
including the room temperature, see Fig. \ref{fig:FIG8-OCS-temps}(c).

Remarkably, the echo amplitude weakly depends on temperature
(in the presented temperature range). This behavior can be understood
with the help of the simplified model introduced in Sec. \ref{sec:Qualitative-discussion}.
The laser pulse is intense enough to induce  efficient phase-space filamentation at all the represented temperatures. With time, the number of filaments grows, and they become thinner. 
As a result, each molecular subgroup forming a single filament has much reduced dispersion of angular velocities (caused by chaotic thermal motion), and it is effectively ``cold''.
Instead of competing with the highly dispersed angular velocities of the initial thermal ensemble, the weak THz pulse competes against the cold narrow filaments. When the change $P_{or}$ of the rotation velocity due to the orienting pulse exceeds the width of the cold filaments, the following orientation echo dynamics becomes insensitive to this width, and, hence, to the initial molecular temperature.  

\begin{figure}
    \includegraphics{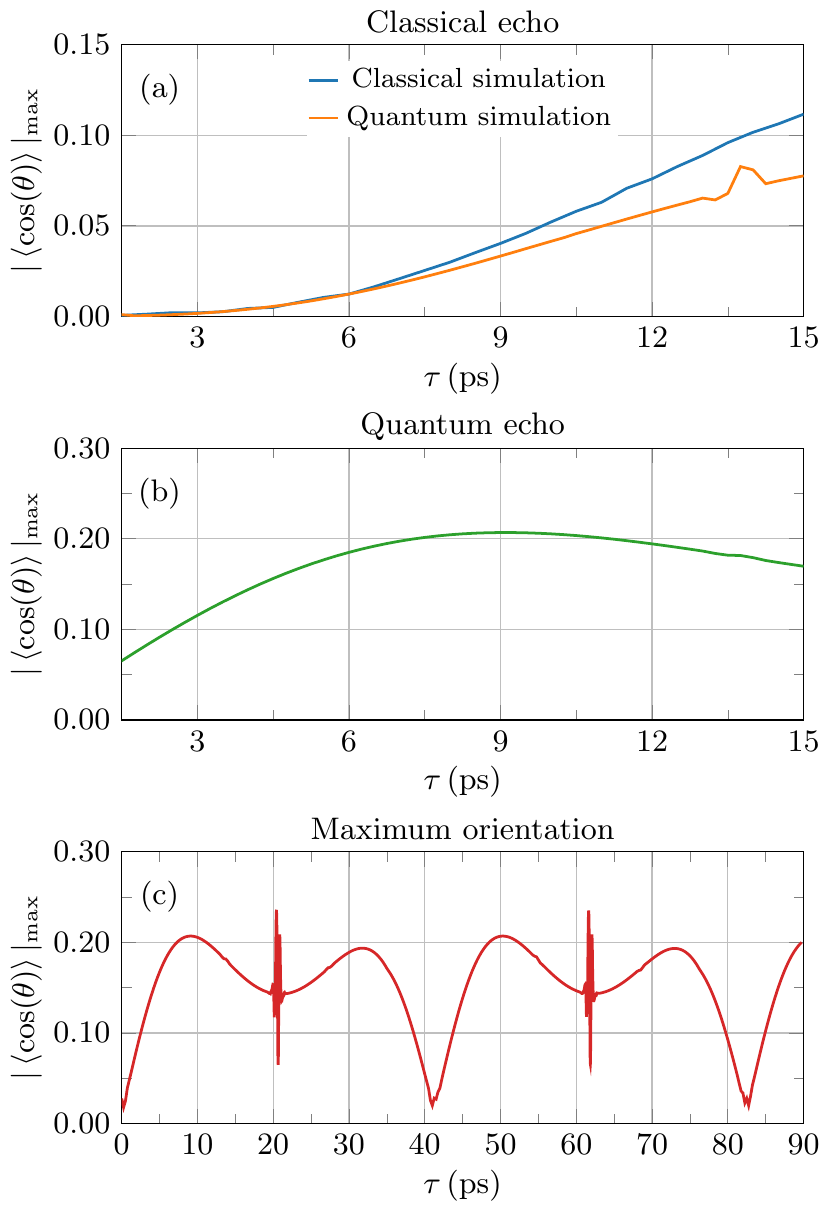} 
    \caption
        {
        Maximum orientation factor (OCS molecule, $T=300\,\mathrm{K}$, absolute value) 
        as a function of $\tau$ (the delay between the prepulse and the orienting THz pulse) 
        in different time windows. 
        (a) During the classical echo emerging after delay equal $2\tau$ after the THz excitation, 
        $t\approx3\tau$.
        (b) During the quantum echo emerging at delay equal $-2\tau$ before the orientation quantum revival, $t\approx T_{rev}-\tau$. 
        (c) The global maximum of the orientation factor after the THz pulse. The field parameters 
        are the same as in Fig. \ref{fig:FIG8-OCS-temps}. For comparison, the maximum orientation 
        factor without the stratifying pulse is $\approx0.0215$.
        \label{fig:FIG9-OCS-time-delay}
        }
\end{figure}

As noted in the caption of Fig. \ref{fig:FIG8-OCS-temps} with the actual THz pulses used here, the dependence of the maximal orientation obtained using the THz pulse only is counterintuitive, namely, the maximal orientation factor slightly increases with temperature instead of decreasing. This is in contrast to the case of impulsive orienting kicks considered in Sec. \ref{sec:Qualitative-discussion}. For relatively long THz pulses, the frequency contents of the pulse match the transitions between higher rotational levels and therefore is more effective in orienting higher temperature ensembles. A similar non-trivial behavior was studied theoretically in \cite{Lapert2012Field}. The dependence of orientation echoes on the pulse duration and other parameters will be elaborated further in a forthcoming publication.

As discussed in Sec. \ref{sec:Qualitative-discussion}, the echo amplitude increases with the time delay between the stratifying and orienting pulses (for fixed temperature and field parameters). Figure \ref{fig:FIG9-OCS-time-delay} shows the maximum orientation during the first classical echo [at delay $2\tau$ after the THz pulse, see panel (a)] and during the quantum echo [at delay $-2\tau$ before the orientation revival, see panel (b)] for relatively short delays $\tau$. For $\tau<15\,\mathrm{ps}$, the magnitude of both types of orientation echoes increases with $\tau$, which is consistent with the qualitative discussion in Sec. \ref{sec:Qualitative-discussion}. Figure \ref{fig:FIG9-OCS-time-delay}(a) shows good correspondence between the quantum and classical results for $\tau<9$. For longer delays, the curves diverge. At the local maximum in Fig. \ref{fig:FIG9-OCS-time-delay}(b) ($\tau\approx9\,\mathrm{ps}$), the enhanced (due to the prepulse) degree of orientation is an order of magnitude higher compared to the case of THz pulse alone.
Figure \ref{fig:FIG9-OCS-time-delay}(c) shows the global orientation maximum for various $\tau$. The dependence on $\tau$ is periodic, and the period equal to half revival, $T_{\mathrm{rev}}/2\approx41.2\,\mathrm{ps}$. When $\tau \approx T_{rev}/4$, the maximum degree of orientation exhibits high-amplitude oscillations. The reason for this high sensitivity is the coalescence of the echo that appears \emph{after} the THz pulse at $t\approx3\tau=3T_{rev}/4$ and the echo that appears \emph{before} the quantum revival at $t\approx T_{rev}-\tau=3T_{rev}/4$. Notice, the orientation can be further enhanced using a stronger prepulse (see Fig. \ref{fig:FIG5-max-ori-afo-Plsr} in Sec. \ref{sec:Qualitative-discussion}).

Previously, related schemes for enhancing the THz- and two-color-induced orientation were applied to cold OCS molecules (at $2\,\mathrm{K}$ in \cite{Egodapitiya2014}, and at $25\,\mathrm{K}$ in \cite{Sonoda2018}). The orientation was substantially enhanced with the help of an aligning laser pulse. The delay between the aligning and orienting pulses was set to $T_{rev}/4$. The authors explained the enhancements using quantum mechanical arguments.
\subsection{Results: symmetric-top molecule}
To demonstrate the robustness of the proposed orientation enhancement
mechanism, we consider an addition example molecule $\mathrm{CH_{3}I}$
that belongs to the class of symmetric-top molecules. The molecular
properties were taken from NIST (DFT, method: CAM-B3LYP/cc-pVTZ-PP)
\cite{johnson1999nist}: $I_{x}=I_{y}\equiv I=68.527\,\mathrm{amu\,\mathring{A}^{2}}$,
$I_{z}=3.23\,\mathrm{amu\,\mathring{A}^{2}}$, $\mu=1.697\,\mathrm{Debye}$,
and $\Delta\alpha=2.951\,\mathrm{\mathring{A}^{3}}$.

\begin{figure}
    \centering
    \includegraphics{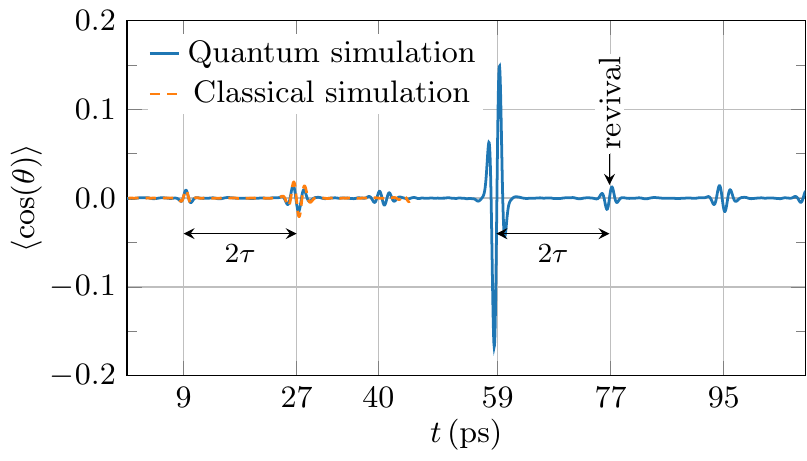} 
    \caption
    {
        Orientation factor for $\mathrm{CH_{3}I}$ molecules calculated quantum mechanically and classically at $T=300\,\mathrm{K}$. The delay between the stratifying and the THz pulses is $\tau=9\,\mathrm{ps}$. The field parameters are the same as in Fig. \ref{fig:FIG8-OCS-temps}, except the THz field intensity $\varepsilon_{0,THz}=0.3\,\mathrm{MV/cm}$. The maximum orientation factor obtained using only the THz pulse is $\approx0.016$.
        \label{fig:FIG10-CH3I-quantum-ori}
    }
\end{figure}

Figure \ref{fig:FIG10-CH3I-quantum-ori} shows the orientation factor
obtained using the quantum simulation at room temperature. The delay
between the pulses was set to $\tau=9\,\mathrm{ps}$ [see Eq. \eqref{eq:electric-field}]. All the
parameters are the same as in Fig. \ref{fig:FIG8-OCS-temps}, except the lower peak amplitude of the THz pulse, $\varepsilon_{0,THz}=0.3\,\mathrm{MV/cm}$. 
On the short time scale, the quantum and classical results are close. The maximum orientation emerges before the quantum revival (during the quantum
echo). The maximum orientation obtained using the THz pulse alone at room temperature is ten times lower, $\approx1.6\%$.

\section{Conclusion \label{sec:Conclusion}}

In this work, we introduced an echo-based approach for enhancing the field-free molecular orientation induced by THz pulses at high temperatures. 
The enhancement mechanism relies on the general nonlinear phenomenon of phase space filamentation and requires a single additional non-resonant laser pulse before the orienting THz excitation. The intense laser pulse effectively eliminates the detrimental thermal broadening by producing multiple separated narrow filaments in the molecular rotational phase space. Molecular subgroups forming individual filaments are effectively cold and easy to orient. The overall significant orientation of the whole ensemble is achieved in the course of echoes forming at set time delays. The various echoes emerging from the combined laser and THz excitation are qualitatively explained using the transparent and intuitive 2D model. In particular, we show that of many echoes possible, the so-called quantum echo, gives rise to the especially pronounced orientation enhancement. 
In the examples presented in the paper, the maximum orientation at room temperature reaches $\sim 20\%$. With prepulses of higher intensity and various optimization procedures, even higher orientation can be expected.  
We demonstrate the robustness of the proposed scheme by using fully quantum simulations of the dynamics of linear (OCS) and symmetric-top ($\rm{CH_3I}$) molecules at room temperature. 
In principle, the temperature range over which the current methodology is operating is not limited to room temperature. Higher temperature regimes could be explored, which may help study materials with low vapor pressure.
The high degree of orientation in a thermal gas paves the way to unprecedented imaging opportunities, chemical reaction control, and more.
\vspace{-3mm}
\begin{acknowledgments}
    L.X. is a recipient of the Sir Charles Clore Postdoctoral Fellowship. 
    This research was made possible in part by the historic generosity 
    of the Harold Perlman Family.
\end{acknowledgments}
\appendix
\section{Impulsive approximation \label{sec:App-A-2D-model-details}}

One of the tools for inducing molecular orientation is a half-cycle
THz pulse \cite{Egodapitiya2014}. The interaction of such a pulse with
the permanent molecular dipole can be modeled using the potential
\begin{equation}
    V_{or}(\theta,t) = -\mu\varepsilon_{or}(t)\cos(\theta),
\end{equation}
where $\mu$ is the molecular dipole moment magnitude, $\varepsilon_{or}(t)$
defines the electric field's time dependence, and $\theta$ is the
polar angle between the molecular axis and the electric field vector. 
For the sake of the model, we assume that $\varepsilon_{or}(t)$ is Gaussian
\begin{equation}
    \varepsilon_{or}(t) = \varepsilon_{0,or}\exp\left[-t^{2}/\sigma_{or}^{2}\right],
\end{equation}
where $\varepsilon_{0,or}$ is the peak amplitude of the electric
field, $\sigma_{or}$ is the width of the pulse. The equation of motion
for the molecular angular momentum $L$ is 
\begin{equation}
    \frac{dL}{dt} = -\frac{dV_{or}(\theta)}{d\theta} 
                  = -\mu\varepsilon_{or}(t)\sin(\theta).
\end{equation}

In the impulsive approximation, when the change in $\theta$ is negligible
during the pulse, the change in $L$ due to the orienting interaction
is 
\begin{equation}
    \Delta L_{or} = -\sqrt{\pi}\mu\sigma_{or}\varepsilon_{0,or}\sin(\theta_{0}),
\end{equation}
where $\theta_{0}$ is the angle at the moment of the pulse, and $P_{or}$
is defined as 
\begin{equation}
    P_{or} = \frac{\mu\sqrt{\pi}}{I}\sigma_{or}\varepsilon_{0,or},
    \label{eq:App-A-P-THz}
\end{equation}
where $I$ is the molecule's moment of inertia.

A standard tool for inducing molecular alignment is non-resonant
femtosecond laser pulses. We use such a pulse as a prepulse 
to stratify the phase space, see Sec. \ref{sec:Qualitative-discussion}. 
The interaction of a molecule with the prepulse is modeled as 
\begin{equation}
    V_{pre}(\theta,t) = -\frac{\Delta\alpha}{4}\varepsilon_{pre}^{2}(t)\cos^{2}(\theta),
    \label{eq:App-A-laser-potential}
\end{equation}
where $\Delta\alpha$ is the polarizability anisotropy of the molecule.
The potential in Eq. \eqref{eq:App-A-laser-potential} is averaged
over the laser's optical cycle. The function $\varepsilon_{pre}(t)$ is the
slowly varying Gaussian envelope of the laser pulse 
\begin{equation}
\varepsilon_{pre}(t) = \varepsilon_{0,pre}\exp\left[-2\ln(2)t^{2}/\sigma_{pre}^{2}\right],
\end{equation}
where $\varepsilon_{0,pre}$ is the peak amplitude of the laser's electric
field, $\sigma_{pre}$ is the full width at half maximum of the laser
pulse intensity. In the impulsive approximation
\begin{equation}
    \Delta L_{pre} = -\sqrt{\frac{\pi}{\ln(16)}}\frac{\Delta\alpha}{4}\sigma_{pre}\varepsilon_{0,pre}\sin(2\theta_{0}),
\end{equation}
and $P_{pre}$ is defined as 
\begin{equation}
    P_{pre} = \sqrt{\frac{\pi}{\ln(16)}}\frac{\Delta\alpha}{4I}\sigma_{pre}\varepsilon_{0,pre}.
    \label{eq:App-A-P-laser}
\end{equation}
\section{Orientation factor \label{sec:App-B-Orientation-factor}}
Here, we derive the formulas for the orientation factor
\begin{equation}
    \braket{\cos(\theta)}(t) 
    = \braket{\mathrm{Re}(e^{i\theta})}(t) 
    = \mathrm{Re}\left[\braket{e^{i\theta}}(t)\right],
\end{equation}
where the angle brackets denote ensemble average,
for two cases:
(i) excitation by an orienting pulse alone, and 
(ii) excitation by a prepulse followed by an orienting kick.
\subsection{Single orienting pulse \label{sec:App-B-single-ori}}
The time-dependent orientation factor after the orienting excitation
is given by the real part of 
\begin{align}
    \braket{e^{i\theta}}(t) & =(2\pi)^{-3/2}\sigma_T^{-1}\nonumber \\
     & \times\int_{-\infty}^{\infty}\int_0^{2\pi}\exp\{i[\theta_0+\omega_0 t-P_{or}\sin(\theta_0)t]\}\nonumber \\
     & \times\exp[-\omega_0^2/(2\sigma_T^2)]\,d\theta_0d\omega_0\nonumber \\
     & =(2\pi)^{-3/2}\sigma_T^{-1}\int_{-\infty}^{\infty}e^{i\omega_0t}\exp[-\omega_0^2/(2\sigma_{T}^{2})]\,d\omega_0\nonumber \\
     & \times\int_0^{2\pi}e^{i\theta_0}\exp\left[-iP_{or}\sin(\theta_0)t\right]\,d\theta_0,
     \label{eq:App-B-ori-integral}
\end{align}
where $\sigma_T=\sqrt{k_B T/I}$, and $\theta(t)=\theta_0+\omega_0t-P_{or}\sin(\theta_0)t$
[see Eq. \eqref{eq:theta-omega-ori}]. Using the Jacobi--Anger
expansion 
\begin{equation}
    e^{iz\sin(\theta)} = \sum_{k=-\infty}^{\infty}e^{ik\theta}J_k(z),
\end{equation}
where $J_{k}(z)$ is the Bessel function of integer order $k$, we
can represent the second integral in Eq. \eqref{eq:App-B-ori-integral}
as 
\begin{align}
    &\int_0^{2\pi}e^{i\theta_0}\exp\left[-iP_{or}\sin(\theta_0)t\right]\,d\theta_0\nonumber \\
    &=\sum_{k=-\infty}^\infty \int_0^{2\pi}e^{i\theta_0}e^{-ik\theta_0}
            J_{k}(P_{or}t)\,d\theta_0\nonumber \\
    &=2\pi J_{1}(P_{or}t),
\end{align}
where all the terms in the sum, except for $k=1$, vanish. Finally,
\begin{equation}
    \braket{\cos(\theta)}(t)=e^{-\sigma_T^2 t^2/2}J_1 (P_{or}t).
    \label{eq:App-B-ori-factor-THz}
\end{equation}
\subsection{Stratifying prepulse followed by orienting kick \label{sec:App-B-both-pre-ori}}
Just after the stratifying kick (prepulse), the rotor's variables
are 
\begin{equation}
    \begin{aligned}
    \theta_1(t) & =\theta_0 + \omega_1t,\\
    \omega_1 & =\omega_0 - P_{pre}\sin(2\theta_0),
    \end{aligned}
    \label{eq:App-B-theta-omega-after-laser}
\end{equation}
After a delay equal to $\tau$, at $t=\tau$, we apply the orienting
kick after which the rotor's variables are 
\begin{equation}
    \begin{aligned}
    \theta_{2}(t) & =\theta_{1}(\tau)+(t-\tau)\omega_{2},\\
    \omega_{2} & =\omega_{1}-P_{or}\sin[\theta_{1}(\tau)].
    \end{aligned}
    \label{eq:App-B-theta-omega-after-THz}
\end{equation}
The orientation factor is given by the real part of 
\begin{align}
    \braket{e^{i\theta}}(t) & =\Theta(t-\tau)\frac{(2\pi)^{-3/2}}{\sigma_T}\int_{-\infty}^{\infty}\int_0^{2\pi}\exp\left[i\theta_2(t)\right] \nonumber \\
     & \times\exp[-\omega_0^2/(2\sigma_T^2)]\,d\theta_0 d\omega_0,
\end{align}
where $\Theta(t-\tau)$ is the step function. We use Eqs. \eqref{eq:App-B-theta-omega-after-laser} 
and \eqref{eq:App-B-theta-omega-after-THz} to express $\theta_2$ in
terms of $\theta_0$ and $\omega_0$. Next, we use the Jacobi--Anger
expansion 
\begin{align}
    \exp&\left\{ iP_{or}(t-\tau)\sin[-\theta_0 - \omega_0 \tau + P_{pre}\sin(2\theta_0)\tau]\right\} \nonumber \\
    &=\sum_{k=-\infty}^{\infty}\left(\exp\left\{ -ik[\theta_0+\omega_{0}\tau-P_{pre}\sin(2\theta_0)\tau]\right\} \right.\nonumber \\
    & \left. \times J_{k}[P_{or}(t-\tau)]\right)
\end{align}
Combining the terms proportional to $\omega_0$ in the exponent,
we get $\exp[i\omega_0t-ik\omega_0\tau]$. Integral $d\omega_0$
then yields 
\begin{align}
    \int_{-\infty}^{\infty}\exp[i\omega_0(t-k\tau)]e^{-\omega_0^2/(2\sigma_T^2)}\,d\omega_0\nonumber \\
    =\sqrt{2\pi}\sigma_Te^{-\sigma_T^2(t-k\tau)^2/2}.
\end{align}
The rest of the terms in the exponent read 
\begin{align}
     & \exp[i\theta_0-iP_{pre}\sin(2\theta_0)t-ik\theta_0+ikP_{pre}\sin(2\theta_0)\tau]\nonumber \\
    = & \exp[i\theta_0(1-k)+iP_{pre}\sin(2\theta_0)(k\tau-t)]\nonumber \\
    = & \sum_{m=-\infty}^{\infty}\exp[i\theta_0(1-k+2m)]J_{m}[P_{pre}(k\tau-t)].
\end{align}
Integral $d\theta_0$ then yields 
\begin{align}
     & \int_{0}^{2\pi}\sum_{m=-\infty}^{\infty}\exp[i\theta_0(1-k+2m)]J_{m}[P_{pre}(k\tau-t)]\,d\theta_0\nonumber \\
     & = 2\pi\sum_{m=-\infty}^{\infty}\delta(1-k+2m)J_{m}[P_{pre}(k\tau-t)]\nonumber \\
     & =
     \begin{cases}
        2\pi J_\frac{k-1}{2}[P_{pre}(k\tau-t)], & \mathrm{odd}\;k,\\
        0, & \mathrm{\mathrm{even\;}}k.
    \end{cases}
\end{align}
Overall, the orientation signal reads 
\begin{align}
    \braket{\cos(\theta)}(t)=\Theta(t-\tau)\sum_{\mathrm{odd}\,k}^{\infty}S_k(t),
    \label{eq:App-B-ori-after-lsr-THz}
\end{align}
where 
\begin{align}
    S_{k}(t) & =e^{-\sigma_{T}^{2}(t-k\tau)^{2}/2}J_{k}[P_{or}(t-\tau)]\nonumber \\
    & \times J_{\frac{k-1}{2}}[P_{pre}(k\tau-t)].
\end{align}
\section{Quantum rigid rotor \label{sec:App-C-Quantum-rigid-rotor}}
The energy eigenstates of the quantum rotor restricted to the $XY$
plane are $\psi_{n}=\exp(i\phi n)/\sqrt{2\pi}$, where $n=0,\pm1,\pm2,\dots$.
The energies (in atomic units, $\hbar=1$) are $E_{n}=n^{2}/(2I)$,
where $I$ is the moment of inertia. Like in the classical case, we
treat the effect of the stratifying (prepulse) and orienting
kicks in the impulsive approximation. The prepulse transforms an
initial wave function $\psi_{-}$ into $\psi_{+}=\exp[iIP_{pre}\cos^{2}(\theta)]\psi_{-}$,
where $P_{pre}$ is given by Eq. \eqref{eq:App-A-P-laser}. Similarly,
for the orienting kick $\psi_{+}=\exp[iIP_{or}\cos(\theta)]\psi_{-}$,
where $P_{or}$ is given by $\eqref{eq:App-A-P-THz}$.

Each of the basis wave functions, $\psi_{n}$ is ``kicked'' and propagated in time. 
The final observable, the ensemble averaged orientation factor, 
$\braket{\cos(\theta)}$ is obtained by thermal averaging over the various 
expectation values $\braket{\Psi_{n}(\theta,t)|\cos(\theta)|\Psi_{n}(\theta,t)}$,
where $\Psi_{n}(\theta,t)$ is the wave function at time $t$ corresponding
to the initial state $\psi_{n}$.

\bibliography{bibliography}

\end{document}